\documentclass[journal]{aa}

\usepackage[dvips]{graphicx}
\usepackage{amsmath}
\usepackage{amssymb}
\usepackage{verbatim}
\usepackage{array}
\usepackage{txfonts}
\usepackage{natbib}

\bibpunct{(}{)}{;}{a}{}{,}

\newcommand{\celltspace}{\rule{0pt}{2.8ex}}
\newcommand{\cellbspace}{\rule[-1.4ex]{0pt}{0pt}}

\begin{document}

\title{New estimates of the gamma-ray line emission\\ of the Cygnus region from INTEGRAL/SPI observations}
\titlerunning{New estimates of the gamma-ray line emission of the Cygnus region}
\author{Pierrick Martin \inst{1,2} \and J\"urgen Kn\"odlseder \inst{1,2} \and Roland Diehl \inst{3} \and Georges Meynet \inst{4}}
\institute{Centre d'Etude Spatiale des Rayonnements, CNRS/UPS, 9, avenue colonel Roche, BP44346, 31028 Toulouse cedex 4, France  \and Universit\'e de Toulouse (UPS), Centre National de la Recherche Scientifique (CNRS), UMR 5187 \and Max Planck Institut f\"ur extraterrestrische Physik, Postfach 1312, 85741 Garching, Germany \and Geneva Observatory, CH-1290 Sauverny, Switzerland}           
\date{Received 25 March 2009 / Accepted 18 June 2009}
\abstract{The Cygnus region harbours a huge complex of massive stars at a distance of 1.0-2.0\,kpc from us. About 170 O stars are distributed over several OB associations, among which the Cyg OB2 cluster is by far the largest with about 100-120 O stars. As a consequence of their successive nuclear-burning episodes, these massive stars inject large quantities of radioactive nuclei into the interstellar medium such as $^{26}$Al and $^{60}$Fe. The gamma-ray line signal from the latter is a solid tracer of ongoing nucleosynthesis.}{We want to compare the decay emission from the Cygnus region with predictions of recently improved stellar models. As a first step, we establish observational constraints upon the gamma-ray line emission from $^{26}$Al and $^{60}$Fe, with particular emphasis placed on separating the emission due to the Cygnus complex from the foreground and background mean Galactic contributions.}{We used the high-resolution gamma-ray spectrometer INTEGRAL/SPI to analyse the $^{26}$Al and $^{60}$Fe decay signal from the Cygnus region. The weak gamma-ray line emissions at 1809 and 1173/1332\,keV have been characterised in terms of photometry, spectrometry, and source morphology.}{The 1809\,keV emission from Cygnus is centred on the position of the Cyg OB2 cluster and has a typical size of 9$^{\circ}$ or more. The total 1809\,keV flux from the Cygnus region is $(6.0 \pm 1.0) \times 10^{-5}$ ph\,cm$^{-2}$\,s$^{-1}$, but the flux really attributable to the Cygnus complex reduces to $(3.9 \pm 1.1) \times 10^{-5}$ ph\,cm$^{-2}$\,s$^{-1}$. The 1809\,keV line centroid is agrees with expectations considering the direction and distance of the Cygnus complex, and the line width is consistent with typical motions of the interstellar medium. No decay emission from $^{60}$Fe has been observed from the Cygnus region and an upper limit of $1.6 \times 10^{-5}$ ph\,cm$^{-2}$\,s$^{-1}$ was derived.}{The determination in a coherent way of the gamma-ray line emission of the clustered OB population of the nearby Cygnus complex will allow a more accurate comparison of observations with theoretical expectations based on the most recent stellar models.}

\maketitle

\section{Introduction}
\label{introduction}

When observing our Galaxy across the electromagnetic spectrum, it is remarkable that at nearly all wavelengths, a strong emission arises from the Cygnus region, while most of the radiation otherwise comes from the inner Galaxy (and from a few close objects such as the Crab and Vela pulsars or the Cassiopeia A remnant). Such a strong broad-band emission points to prodigious amounts of mass and energy release in that direction, at intermediate distances from us. Nevertheless, the true nature, scale, and extent of the phenomenon underlying such a powerful radiative output has mostly remained unclear for a long time.\\
\indent The line of sight to the Cygnus region (assumed in the following to be the part of the Galactic plane located between longitudes 70$^{\circ}$ and 90$^{\circ}$) is tangential to the Local spiral arm. In that direction we are viewing material located close to the solar circle, which strongly compromises a solid evaluation of distances between 0 and 4-5\,kpc from kinematic arguments\footnote{This comes from the motions of distant objects along our line of sight, due to Galactic rotation, being lower than the typical proper motions of interstellar clouds and star-forming regions (especially when massive star clusters are present, which is the case here)} \citep{Schneider:2006}. Strong absorption towards the Cygnus area (up to 32 visual magnitudes) has been another hurdle on the way to understanding the layout of the region \citep[see Fig. 1 of][]{Schneider:2006}. As a consequence, it has long been debated whether the Cygnus region results from a superposition of several objects/structures distributed along the line of sight between 0.5 and 3.5\,kpc \citep{Bochkarev:1985,Wendker:1991,Uyaniker:2001}, or whether it is a single physical entity located around 1.5\,kpc and likely produced by an unusual concentration of stars \citep{Veron:1965,Cash:1980}. The latter scenario was claimed to be proven by the HEAO-1 and ROSAT observations that revealed a large soft X-ray ring in the Cygnus direction, which could result from a gigantic superbubble located at $\sim$ 2\,kpc from us, behind the Great Cygnus Rift, and energised by a series of 30-100 supernova explosions \citep{Cash:1980}.\\
\indent \citet{Humphreys:1978} originally inventoried 9 OB associations in the Cygnus region, but subsequent works have cast doubt on the reality of Cyg OB4, 5, and 6. At present, the clustered stellar content of the region is thought to consist of Cyg OB1, 2, 3, 7, 8, and 9 plus a dozen open clusters, most of which may well be related to Cyg OB1 and OB3 \citep{Knodlseder:2002,Le-Duigou:2002}. Most of these stellar groups are located between 1 and 2\,kpc from us, with typical distance uncertainties of 0.3-0.5\,kpc. Recently, a stellar census based on the 2MASS catalogue has revealed that the richness of the Cyg OB2 star cluster actually exceeds the previous estimates by an order of magnitude. Indeed, Cyg OB2 contains about 100-120 O stars and could well be the first example of a young Galactic globular cluster \citep{Knodlseder:2000,Comeron:2000}. While this does not constitute a definitive answer to the question of the origin of the strong emission from Cygnus, it seems obvious that such a titanic object will generate a considerable energy output and strongly influences the surrounding interstellar medium. In particular, \citet{Lozinskaya:2002} have shown that Cyg OB2, with its increased membership, could alone, through its collective stellar winds, account for the X-ray superbubble and its associated system of giant optical filaments.\\
\indent Another argument supporting the idea of a single physical entity has been provided by the work of \citet{Schneider:2006}. From KOSMA sub-millimetre and MSX mid-infrared observations, the authors have shown that the molecular material in Cygnus belongs to two huge complexes of clouds (with masses of 2.8 and 4.8 $\times$ 10$^{5}$\,M$_{\odot}$), which are very likely the remainder of a giant molecular cloud out of which Cyg OB2, OB9 and possibly OB1 formed. This mass distribution is now influenced by the UV light and winds from these stellar associations and is a site of active star formation. This coherent structure lies at $\sim$ 1.5\,kpc, the approximate distance to the Cyg OB2 cluster.\\
\indent Although the relation of Cyg OB3, OB7, and OB8 to this structure has not been established, we refer in the following to the 6 OB associations and associated open clusters of the Cygnus region as the "Cygnus complex". With this term, we aim at gathering all these rich and nearby stellar groups that are relatively correlated in terms of distance (between 1 and 2\,kpc) and that very likely contribute to the specificity of the Cygnus region. Moreover, we note that the current uncertainties on the ages and distances of these stellar clusters do not allow firm exclusion of the scenario of a connexion between them all.\\ 
\indent The Cygnus complex amounts to a total of $\sim$ 170 O stars, among which a dozen WR stars \citep{Knodlseder:2002}. The region is therefore quite young, and this is further proven by the lack of radio supernova remnants (SNRs) or pulsars. Indeed, the Green catalogue\footnote{Green D.A., 2006, "A Catalogue of Galactic Supernova Remnants (2006 April version)", Astrophysics Group, Cavendish Laboratory, Cambridge, UK (http://www.mrao.cam.ac.uk/surveys/snrs/).} gives about 10 SNRs between 70$^{\circ}$ and 90$^{\circ}$, but according to a compilation of distances by \citet{Kaplan:2004} none of these objects can be related to an association of the Cygnus complex. The closest object to the Cygnus complex is $\gamma$-Cygni (SNR 078.2+2.1), located at 1.2\,kpc from us. From the sensitivity of their radio survey of the region, \citet{Wendker:1991} concluded that no other SNR exists up to a distance of 10\,kpc. Additional evidence for the relative youth of the complex comes from the absence of pulsars towards Cygnus up to 4\,kpc, as indicated by the ATNF catalogue\footnote{http://www.atnf.csiro.au/research/pulsar/psrcat/} \citep{Manchester:2005}. These arguments also seem to contradict the scenario considered by \citet{Cash:1980} in which 30-100 supernovae over the past few Myr would have blown the Cygnus superbubble \citep[one should note, however, that the SNR ages are typically much lower than a Myr, which could explain the present lack; see][]{Frail:1994}.\\
\indent The Cygnus region therefore constitutes a promising opportunity to look deeper into the physics of massive stars, in particular the nucleosynthetic processes that enrich the interstellar medium (ISM) with new elements. In the work presented here, we study the nucleosynthetic activity in the Cygnus region and its complex of massive stars, as traced by the gamma-ray line emission from the decay of $^{26}$Al and $^{60}$Fe. These two radio-isotopes are thought to be produced predominantly by massive stars at various stages of their evolution, from the main sequence to the supernova explosion \citep[see][for a review of the different production sites]{Limongi:2006}. The release of $^{26}$Al in the ISM occurs via stellar winds (of Wolf-Rayet stars mainly) and supernova explosions, while $^{60}$Fe is only ejected by supernova explosions. With lifetimes of 1 and 2\,Myr respectively, $^{26}$Al and $^{60}$Fe accumulate in the ISM and their decay signal builds up to levels that are within reach of gamma-ray instruments. The 1809\,keV emission of $^{26}$Al from the whole Galaxy has been imaged by CGRO/COMPTEL and INTEGRAL/SPI \citep{Pluschke:2001,Diehl:2006,Wang:2008} and shows a strong emission from the Cygnus region; the 1173 and 1332\,keV Galactic emission from $^{60}$Fe has been observed by RHESSI and INTEGRAL/SPI \citep{Smith:2004,Harris:2005,Wang:2007} but no emission from Cygnus has been recorded, which indicates that none or only a few supernovae went off and did release $^{60}$Fe in that region.\\
\indent Because of its strong 1809\,keV decay emission, the Cygnus region was soon identified as an interesting target for studying the nucleosynthesis of massive stars. Indeed, the Galactic 1809\,keV emission results from a situation involving many uncertain parameters: the distribution of massive stars in the Galactic disk and bulge, the Galactic metallicity profile, the star formation rate,...etc. Moreover, there may well be secondary sources of $^{26}$Al and $^{60}$Fe, like the supernovae of type Ia, the novae, or the AGB stars, with different timescales for the release of both isotopes. All these elements hinder accurate determination of the expected decay fluxes to which the observations could be compared. In contrast, the Cygnus region and its complex of massive stars form a restricted, more homogeneous population, the characteristics of which we know better. In addition, whereas the Galaxy is in a steady-state situation regarding its 1809\,keV emission, the Cygnus complex is a young, evolving population, which thus allows assessment of not only the total yields of $^{26}$Al and $^{60}$Fe but also the history of their ejection in the ISM.\\
\indent Since the cartography of the Galactic 1809\,keV emission was achieved by CGRO/COMPTEL, a number of studies have been devoted to the nucleosynthesis of $^{26}$Al and $^{60}$Fe in the Cygnus region \citep[see for example][]{Del-Rio:1996,Pluschke:2000,Cervino:2000,Pluschke:2002,Knodlseder:2002}. The latest studies agree that the observed level of the 1809\,keV emission from Cygnus is a factor of 2-3 above the predictions, when the latter are based on the theoretical yields by \citet{Meynet:1997} and \citet{Woosley:1995} and on the proven stellar content of the Cygnus region (including the upward revision of the Cyg OB2 membership). \citet{Pluschke:2000} shows that the discrepancy could be alleviated if one includes the enhanced (but highly uncertain) yields from massive close binary systems in the calculation. Alternatively, \citet{Pluschke:2002} argues that the problem may well come from underestimating the stellar content of the Cygnus OB associations, because of a strong interstellar extinction, as was proven to be the case for Cyg OB2. In \citet{Knodlseder:2002}, the authors eventually speculate that future stellar models, especially models including the effects of stellar rotation, could solve the dilemma.\\
\indent In recent years, the models of massive stars have indeed undergone major improvements, such as the introduction of stellar rotation or the use of revised mass-loss rates or nuclear cross-sections \citep[see for example][]{Meynet:2000,Meynet:2003}. The effects of metallicity have also been explored by \citet{Meynet:2005}, while \citet{Limongi:2006} computed the contributions of the supernovae of type Ib/c in a consistent way. Meanwhile, the launch in 2002 of the INTEGRAL gamma-ray observatory has provided the community with a new instrument for addressing the problem from the observational side. In particular, INTEGRAL has devoted substantial observing time to the Cygnus region through dedicated programmes, and its high-resolution spectrometer SPI should allow the previous photometric measurements to be confirmed and add spectrometric information about the gamma-ray decay lines.\\
\indent The aim of our work is to re-evaluate the situation of the gamma-ray line emission from Cygnus, in terms of observations, as well as theoretical predictions. For the present paper, we used the INTEGRAL/SPI observations to characterise in detail the $^{26}$Al and $^{60}$Fe decay emission from the Cygnus region. The analysis of the SPI data is presented in Sect. \ref{analysis} and the results will be exposed in Sects. \ref{cygnus} and \ref{galaxy}. The observational constraints thus obtained will then be compared in a forthcoming paper to theoretical predictions based on the most recent models of massive stars' nucleosynthesis.

\section{Data analysis}
\label{analysis}

The SPI spectrometer is a coded-mask telescope with a large field of view of $16^{\circ}\times16^{\circ}$ and an angular resolution of 2.8$^{\circ}$ \citep[detailed information about the SPI instrument can be found in][]{Vedrenne:2003,Roques:2003}. And yet, the complexity of the response function means that reconstructing a given source intensity distribution cannot be achieved from a single exposure using standard correlation methods. Instead, we have to resort to multiple exposures (following a \textit{dithering pattern}) and then to model-fitting approaches.\\
\indent Thanks to an array of 19 high-purity germanium detectors, SPI performs spectrometry of gamma radiations with energies between 20 keV and 8 MeV, with a spectral resolution of about 2 keV FWHM at 1 MeV. Interactions of gamma photons with the detectors can be of two types: \textit{single events} (SE) when the incoming photon deposits energy in one detector only, and \textit{multiple events} (ME) when the incoming photon deposits energy in several adjacent detectors through Compton diffusion and pair creation. In the following, we only make use of SE data. It should be noted that detector 2 and 17 failed on revolutions 142 and 210.\\
\indent The data produced by SPI are strongly background-dominated (signal-to-noise ratio of less than 1\%) because substantial hadronic interactions and electromagnetic cascades are induced by solar and cosmic-ray particles hitting the satellite. These result in an instrumental background composed of a continuum and a certain number of lines of various intensities, some of which correspond to the same nuclear transitions we are searching for (for example the line at 1809\,keV from the de-excitation of $^{26}$Mg inside the instrument).\\
\indent Extracting the weak gamma-ray line signals from astrophysical objects therefore requires an accurate modelling of that background noise; in our work, this is achieved through combinations of \textit{activity tracers} that are hypothesised to reproduce the various trends and time evolutions of the background. In this effort, the most utilised tracer is the so-called GEDSAT, which corresponds to the rate of saturating events in the detectors (energy deposits above 8\,MeV). The GEDSAT has indeed proved to be strongly correlated to the SPI data over a broad energy range; in addition, its count rate ($\sim$200\,cts\,s$^{-1}$) is about two orders of magnitude higher than the typical count rate measured at gamma-ray line energies, which means that it suffers little statistical uncertainty and can readily be used to draw predictions. Nevertheless, the GEDSAT is not a perfect tracer and it needs to be complemented and adjusted, as we see below. More information about the instrumental background and our methodology for analysing the SPI data can be found in \citet{Jean:2003} and \citet{Martin:2009}, respectively. 

\subsection{Data set}

The data used for this analysis are all-sky data that were collected between revolutions 7 and 494 of the INTEGRAL satellite (the duration of one INTEGRAL orbit is $\sim$3 days). A filtering was applied to exclude all pointings during which abnormally high count rates (due to solar flares and periodic radiation belt crossings) occurred, so they preclude any detection of celestial signals. The total effective observation time eventually amounts to 63.2\,Ms, among which 10.8\,Ms cover the Cygnus region and 11.2\,Ms are high-latitude pointings that we use as empty fields. All SE data with energies in the 1780-1830, 1150-1190, and 1300-1350\,keV ranges were selected and binned into 1\,keV-wide bins.

\subsection{Background models}

The physics underlying the instrumental background of the SPI spectrometer is complex and not fully elucidated despite considerable efforts \citep[see for example][]{Weidenspointner:2003}. The development of background models based on \textit{activity tracers} therefore proceeds through trial-and-error and there is no general prescription that works successfully over the entire energy range of the instrument (reflecting that the origins of the background are specific to each energy domain).\\
\indent When scaled to the mean level of the data, these background models reproduce the time evolution of the instrumental background noise reasonably well. However, as we are looking for gamma-ray signals that are less than 1\% of the data, such a global scaling is not accurate enough and the background models need to be finely adjusted.\\
\indent In the present work, we study three different gamma-ray lines and it turned out that each energy band required a specific model:
\begin{enumerate}
\item 1809\,keV line: the background model consists of two components, one to account for the background continuum and one for the background lines. The time evolution of each component was assumed to be proportional to the GEDSAT rate in each detector, and the level of the background continuum was taken from two reference energy bands at 1786-1802 and 1815-1828\,keV. The continuum component was adjusted over 1 day intervals while the line component was adjusted over 100 days intervals.
\item 1173\,keV line: the background model consists of three components, one for the background continuum and two for the background lines. The time variability of the background continuum was assumed to follow the GEDSAT rate and the level of the background continuum was taken from a reference energy band at 1178-1182\,keV. The evolution of the background lines intensity is given by the GEDSAT added to a linear increase over time (likely due to the build-up of medium-lived $^{60}$Co in the instrument as a result of continuous activation). The continuum component was adjusted over 1 day intervals while the line components were adjusted over three time periods limited by the failures of detectors 2 and 17.
\item 1332\,keV line: the background model consists of a single component proportional to the GEDSAT rate and adjusted over 3 days intervals. In this energy range, the variability of the background noise was stronger and thus required to adjust the model on shorter timescales, which decreased the sensitivity of the instrument to the weak celestial signal. As a consequence, the 1332\,keV line measurement only marginally contributes to the $^{60}$Fe results.
\end{enumerate}
The background models presented above are fitted to the data, with the proper parameter set, simultaneously to models for the celestial contribution.

\subsection{Model fitting}

As already mentioned, the SPI data cannot be inverted and the astrophysical signals are extracted by fitting models of intensity distributions to the data. This can be done with iterative methods such as Richardson-Lucy or Maximum Entropy that reconstruct the sky intensity distribution pixel by pixel, by setting the flux in each sky element to the value that best accounts for the data (in a statistical sense). An alternative is to set the shape of the source model (from observations at other wavelengths for instance) and to adjust its overall flux. The latter method is the one we used because it ensures that the result will be astrophysically meaningful and compatible with the performances of the instrument (on angular resolution especially), while reconstruction algorithms often generate artefacts from statistical fluctuations and uneven exposure.\\
\indent In the present work, simple analytical models, such as two-dimensional Gaussian profiles and a ring model for the Galaxy, were used for the sky contributions (the details of these models will be given below). Each of these intensity distributions is mapped into the data space through a convolution with the instrument response function (IRF) for each pointing of the data set and then forms a sky model component. The sky model components and the background models are then fitted simultaneously to the data using a Maximum Likelihood criterion for Poissonian statistics.\\
\indent Once all models (sky and background) were fitted to the data, the quality of the analysis was assessed by examining the residuals (data counts minus fitted background and sky models counts). Since we are working with background-dominated data, the modelling of the instrumental background can easily introduce systematic errors that might be compensated for by artificial signals from the sky. In addition, the assumption made on the sky intensity distribution might well be inappropriate or incomplete, which would introduce bias or leave unaccounted for events in the data space.\\
\indent To evaluate the quality of our analysis, we studied the distribution of the residuals  after back-projection on the sky. Basically, for all pointings of the data set, the residuals in each detector were uniformly projected across the mask (or more precisely "across" the instrument response function) over the sky region intersected by the field of view. The distribution of these sky residuals should then follow a statistical distribution, the shape of which is derived from simulated observations. This method is described in greater detail in \citet{Martin:2009} and proved that our hypotheses regarding both the background noise and the sky intensity distribution are satisfactory and that the results obtained are not affected by systematic errors.

\section{The gamma-ray line emission from Cygnus}
\label{cygnus}

\subsection{Morphology of the 1809\,keV emission}
\label{cygnus_morpho}

We have characterised the morphology of the 1809\,keV emission from the Cygnus region through a model-fitting approach. Different models, in terms of size and position over the Cygnus region, were fitted to the data (simultaneously to a model for the instrumental background), and after a correction for the number of independent trials, the model with the highest likelihood criterion was selected. This method has been tested on simulated observations with various true source sizes and positions and proved successful.\\
\indent Due to small statistics on the astrophysical fluxes and poor angular resolution of the instrument, the morphology of the Cygnus 1809\,keV emission cannot be determined with a high degree of accuracy. Therefore, we focused on determining of the extent and centre of the 1809\,keV emission. We considered 6 model types: point-source and 2D Gaussians with standard deviations $\sigma$ of 1$^{\circ}$, 2$^{\circ}$, 3$^{\circ}$, 4$^{\circ}$, and 5$^{\circ}$ (the typical size of these Gaussian models is given by $\sim 3\sigma$). Each of these 6 models was moved over a grid of 176 positions ranging from 65$^{\circ}$ to 95$^{\circ}$ in longitude and -10$^{\circ}$ to 10$^{\circ}$ in latitude, with 2$^{\circ}$ steps in each direction.\\
\indent From this analysis, the best-fitting model is a $3^{\circ} \times 3^{\circ}$ Gaussian centred on $(l,b)=(81^{\circ},-1^{\circ})$. Typical uncertainties are $\sim 3^{\circ}$ on the size of the model and $\sim 3^{\circ}$ on the position (in both longitude and latitude). However, the maximal extent of the diffuse emission at 1809\,keV from Cygnus is poorly constrained because of intrinsic limitations of the coded-mask imaging system. Indeed, a coded-mask instrument is unable to detect a flat intensity distribution because the latter leads to a uniform flux over the detector plane, which cannot be distinguished from instrumental background. Flat or very extended intensity distributions can only be detected by alternating pointings on the source (ON pointings) with pointings to empty fields (OFF pointings), in which case the coded-mask instrument is used as a mere collimator. For the Cygnus extended emission at 1809\,keV, it turned out that sources with typical sizes above 8-10$^{\circ}$ cannot be distinguished (provided most of the emission remains within the field of view); therefore, our best-fitting model should formally be considered as a lower limit. It should be noted, however, that the typical size of the Cygnus 1809\,keV emission on the COMPTEL map is consistent with our best-fitting model.

\subsection{Characteristics of the 1809\,keV emission}
\label{cygnus_spectro}

\begin{figure}[t]
\begin{center}
\includegraphics[width=\columnwidth]{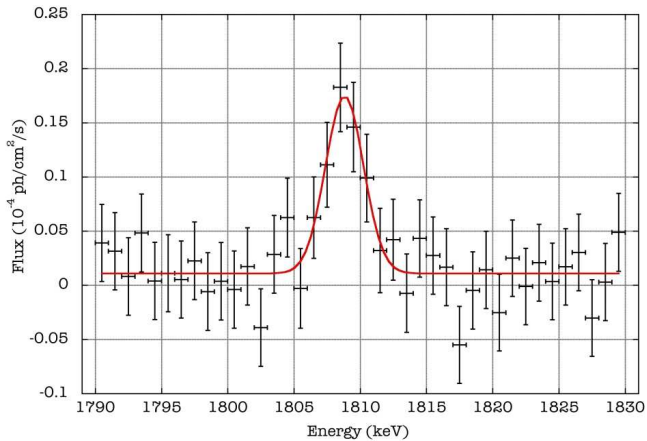}
\caption{Spectrum of the 1809\,keV emission from the Cygnus region, from about 4 years of INTEGRAL/SPI observations. The red line represents the best Gaussian fit to the data points.}
\label{spec26Al_cygnus}
\end{center}
\end{figure}

By representing the 1809\,keV emission from the Cygnus region by the $3^{\circ} \times 3^{\circ}$ Gaussian model determined in \ref{cygnus_morpho}, the fit to the data at 1780-1830\,keV yields the spectrum of Fig. \ref{spec26Al_cygnus}. The centroid of the line is 1808.8 $\pm$ 0.4\,keV, which is consistent with the laboratory value of 1808.6\,keV and hence indicates that the emitting medium has no or only a small bulk motion with respect to the observer. The width of the line is 3.4 $\pm$ 1.0\,keV (FWHM) and is consistent with the 3.0\,keV (FWHM) spectral resolution of the SPI spectrometer. Considering the statistical uncertainty on the line width, the emitting medium may be affected by expansion or turbulent velocities with values up to $\sim$ 200\,km\,s$^{-1}$.\\
\indent Based on our Gaussian model, the flux from the Cygnus region in the 1806-1812\,keV band is $(6.0 \pm 1.0) \times 10^{-5}$ ph\,cm$^{-2}$\,s$^{-1}$. This value, however, gives the total flux seen towards the Cygnus region and thus includes the contribution from the Cygnus complex located at about 1.0-2.0 kpc from the Sun, plus foreground and background contributions from the Galaxy as a whole. If a model for the Galactic 1809\,keV emission (to be described in \ref{galaxy_alu}) is fitted to the all-sky data simultaneously to our Gaussian model, the flux associated with the Gaussian model falls to $(3.9 \pm 1.1) \times 10^{-5}$ ph\,cm$^{-2}$\,s$^{-1}$. We interpret the latter flux as being mostly caused by the OB associations and open clusters of the Cygnus complex, while the remaining $\sim 2.0 \times 10^{-5}$ ph\,cm$^{-2}$\,s$^{-1}$ are very likely due to a more diffuse population, distributed all along the line of sight. This population may include, for instance, unclustered WR and OB stars (runaway stars or B stars from previous star formation episodes, see \citet{Comeron:2007} and \citet{Comeron:2008}), or isolated SNRs.

\subsection{Search for 1173/1332\,keV emission}
\label{cygnus_iron}

\begin{figure}[!t]
\begin{center}
\includegraphics[width=\columnwidth,height=6.44cm]{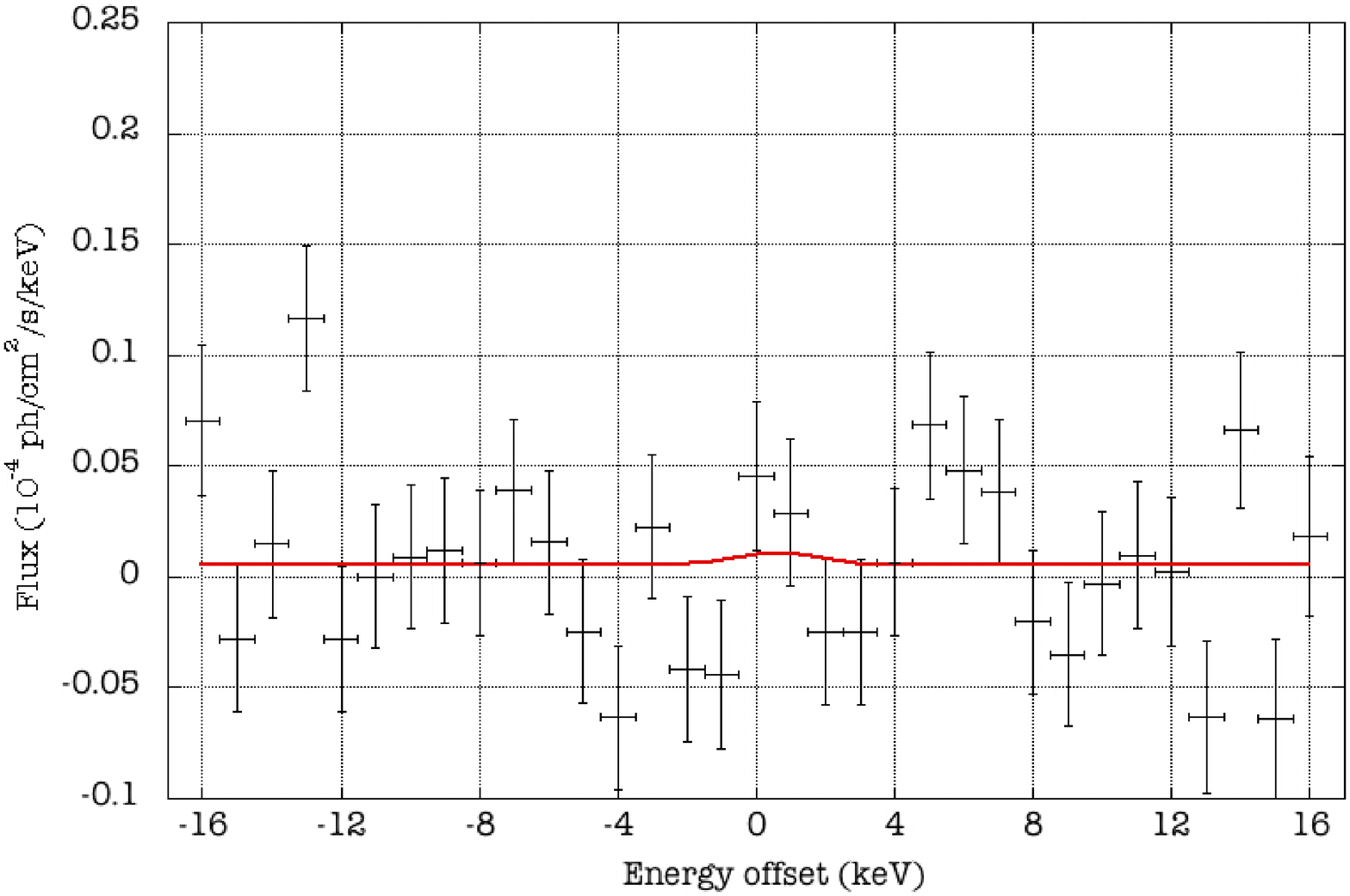}
\caption{Spectrum of the combined 1173/1332\,keV emission from the Cygnus region, from about 4 years of INTEGRAL/SPI observations. The red line represents the best Gaussian fit to the data points.}
\label{spec60Fe_cygnus}
\end{center}
\end{figure}

As is shown in \ref{galaxy_iron}, the Galactic $^{60}$Fe decay emission is about 7 times weaker than the Galactic $^{26}$Al decay emission. The 1809\,keV line allows morphology studies, either over the Galaxy or more specifically in the Cygnus region, which cannot be envisaged with the $^{60}$Fe lines. In the case of the Cygnus region, a study of the morphology of the 1173/1332\,keV emission may be impossible simply because there may not be any 1173/1332\,keV signal at all. Indeed, since the region seems quite young, it may well be that none or few supernova occurred  and thus little $^{60}$Fe has been released in the ISM, if any.\\
\indent Using the same model for the Cygnus region as in \ref{cygnus_spectro} and fitting this model to the data at 1150-1190 and 1300-1350\,keV gives the spectrum of Fig. \ref{spec60Fe_cygnus}. No signal shows up and an upper limit on the flux of $1.6 \times 10^{-5}$ ph\,cm$^{-2}$\,s$^{-1}$ can be derived (2$\sigma$ upper limit assuming the line is not broadened, as is the case at 1809\,keV). We emphasise here that this upper limit applies to the Cygnus region and is therefore conservative if used for the Cygnus complex.

\subsection{Discussion}
\label{cygnus_discu}

The characteristics of the gamma-ray line emission from Cygnus agree with the current picture of the region as derived from observations at other wavelengths.\\
\indent First of all, the absence of $^{60}$Fe decay signal from that direction suggests that no or very few supernovae went off in the Cygnus complex, which is compatible with the lack of SNRs or radio pulsars. Then, the extended emission from $^{26}$Al decay is centred on a position which is consistent with the position of the Cyg OB2 cluster; the latter is expected to dominate the energetics and nucleosynthesis of the complex given its richness compared to the other OB associations. The negligible shift in the 1809\,keV line is in accordance with the expected relative motion due to Galactic rotation for the direction and mean distance of the Cygnus complex, and the width of the line is consistent with typical ISM motions.\\
\indent The extent of the 1809\,keV emission reflects the diffusion of the ejected $^{26}$Al in the ISM; in particular, the $^{26}$Al is expected to fill to some extent the superbubbles blown by the various OB associations. Inside these low-density cavities, the radioisotope is affected by expansion motions that accompany the growth of the superbubble and may cause some broadening of the decay line.\\
\indent Our estimate of the total 1809\,keV flux from the Cygnus region is $(6.0 \pm 1.0) \times 10^{-5}$ ph\,cm$^{-2}$\,s$^{-1}$, in agreement with previous measurements. The flux values obtained with COMPTEL were in the range 7-8 $\times 10^{-5}$ ph\,cm$^{-2}$\,s$^{-1}$, depending on the integration region and imaging method used, with a typical uncertainty of about 1.5 $\times 10^{-5}$ ph\,cm$^{-2}$\,s$^{-1}$ \citep[see for instance][]{Del-Rio:1996,Pluschke:2001a}. Other measurements from SPI are in the range 6-7 $\times 10^{-5}$ ph\,cm$^{-2}$\,s$^{-1}$, here again depending on the imaging method and integration region, with a typical uncertainty of about 1.2 $\times 10^{-5}$ ph\,cm$^{-2}$\,s$^{-1}$ \citep{Wang:2007b}.\\
\indent The most interesting result of our study is that analysing the Cygnus region within a Galactic context has revealed that the 1809\,keV flux really attributable to the Cygnus complex of massive stars is only $\sim$ 65\% of the total flux coming from that direction. This point is essential to the comparison of observed fluxes with the theoretical predictions, because the latter are often mostly based on the clustered stellar content, which is better inventoried than the isolated or non-clustered sources. It is interesting to note here that \citet{Pluschke:2002} find that the contribution of non-clustered sources may amount to about 20-30\%, which tallies with our observational estimate; however, their result was based on an assumed correction of the stellar content of the Cygnus OB associations that has never been observationally substantiated afterward.\\
\indent Last of all, we want to emphasise that our result is based on the assumption of a Gaussian profile for the emission of the Cygnus complex, whereas the true intensity distribution may well be more structured. As already said, the SPI (or even COMPTEL) data do not allow accurate determination of the morphology of the diffuse 1809\,keV emission from Cygnus and, as such, our assumption of a Gaussian profile for the extended source is a potential source of error.

\section{The gamma-ray line emission from the Galaxy}
\label{galaxy}

\subsection{The 1809\,keV Galactic emission}
\label{galaxy_alu}

\begin{figure}[t]
\begin{center}
\includegraphics[width=\columnwidth]{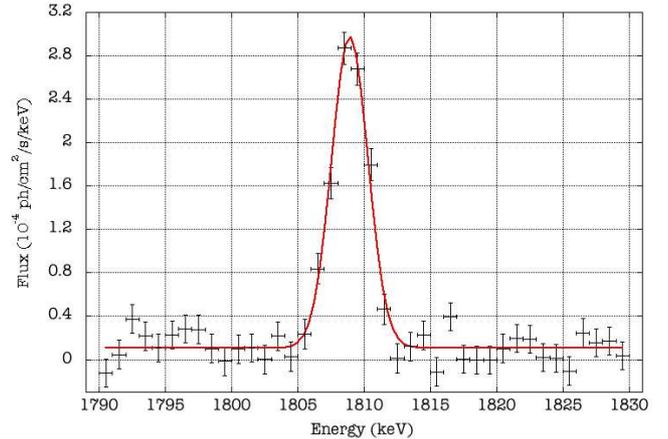}
\caption{Spectrum of the 1809\,keV emission from the Galaxy (Sco-Cen and Cygnus regions excluded), from about 4 years of INTEGRAL/SPI observations. The red line represents the best Gaussian fit to the data points.}
\label{spec26Al_galaxy}
\end{center}
\end{figure}

The detailed study of the Cygnus gamma-ray line emission required to properly take the mean Galactic emission into account in that direction. Therefore, the following presents the model used for the mean Galactic emission at 1809\,keV.\\
\indent When studying the $^{26}$Al Galactic emission, the most important quantities to be determined are the total mass of $^{26}$Al and its distribution throughout the Galaxy; and yet, as already stated, the small statistics on the astrophysical fluxes and the poor angular resolution of the instrument severely restrict the information that can be derived from the SPI observations. The main source of Galactic $^{26}$Al are the massive stars, the vast majority of which clusters in the spiral arms, but our data do not have the degree of accuracy necessary for unambiguous reconstruction of a sophisticated spiral arm pattern from the intensity distribution observed on Earth. Instead, we focus on global quantities such as the azimuthally-averaged radial distribution of $^{26}$Al in the Galaxy.\\
\indent We used a five-component model in which the Galaxy is divided into five concentric rings, each ring containing a certain mass of $^{26}$Al distributed as
\begin{equation}
\label{eq_ringdistrib}
\rho(R_{i} \leq R \leq R_{i+1}) = K_{i} \times e^{-\frac{z}{z_0}} \qquad (i= 1, ..., 5)
\end{equation}
where $R$ is the galactocentric radius, the $R_i$ values define the edges of the rings, and the $K_i$ are the scaling factors set by the model-fitting process. The distribution of the radio-isotope is assumed to be uniform radially inside each ring and exponential vertically, with a characteristic height $z_0$. We started our analysis with a value of $z_0=90$\,pc, which corresponds to the characteristic thickness (within the solar circle) of the molecular gas disk out of which massive stars form \citep{Bronfman:2000}. This parameter will be discussed in more detail later on.\\
\begin{table}[b]
\caption{Galactic mass distribution of $^{26}$Al obtained from our analysis of 4 years of INTEGRAL/SPI observations (see text).}
\begin{center}
\begin{tabular}{|c|c|}
\hline
\celltspace \cellbspace ring 1 (0-3\,kpc) &  0.17 $\pm$0.07\,M$_{\odot}$\\
\hline
\celltspace \cellbspace ring 2 (3-6\,kpc) &  1.18 $\pm$0.15\,M$_{\odot}$\\
\hline
\celltspace \cellbspace ring 3 (6-8\,kpc) &  0.18 $\pm$0.08\,M$_{\odot}$\\
\hline
\celltspace \cellbspace ring 4 (8-9\,kpc) &  0.15 $\pm$0.04\,M$_{\odot}$\\
\hline
\celltspace \cellbspace ring 5 (9-12\,kpc) &  -0.28 $\pm$0.18\,M$_{\odot}$\\
\hline
\celltspace \cellbspace Total $^{26}$Al mass &  1.68 $\pm$0.19\,M$_{\odot}$\\
\hline
\end{tabular}
\end{center}
\label{tab_26Alringmass}
\end{table}
\indent This five-component model turned out to be insufficient to account for the data and should be supplemented by two components: the Gaussian model determined in \ref{cygnus_morpho} for the Cygnus diffuse emission and another Gaussian model for a very extended emission from the Sco-Cen region. The Sco-Cen emission was found to be represented well by a $5^{\circ} \times 5^{\circ}$ Gaussian model centred on $(l,b)=(0^{\circ},30^{\circ})$, which is nearly consistent in position and extent with the superbubble blown by the Upper-Scorpius OB association \citep[see][]{de-Geus:1992}.\\
\indent Fitting these seven intensity distributions (five Galactic rings + Cygnus + Sco-Cen) to the data (simultaneously to a model for the instrumental background) gives the $^{26}$Al mass distribution summarised in Table \ref{tab_26Alringmass} (Note that ring 5 is not used in the computation of the total $^{26}$Al mass because of its non-physical value and large uncertainty). The total corresponding flux from the Galactic plane in the 1806-1812\,keV band amounts to (8.4 $\pm$ 1.7) $\times$ 10$^{-4}$\,ph\,cm$^{-2}$\,s$^{-1}$. In addition to this azimuthally-averaged radial distribution of $^{26}$Al in the Galaxy, the nearby regions of Cygnus and Sco-Cen bring additional emission at a level of (3.9 $\pm$ 1.1) and (6.2 $\pm$ 1.6) $\times$ 10$^{-5}$\,ph\,cm$^{-2}$\,s$^{-1}$ respectively. This paper focuses on the Cygnus region and the 1809\,keV emission from Sco-Cen will be discussed in a forthcoming paper \citep{Diehl:2009}.\\
\indent The spectrum of the 1809\,keV emission from the Galactic disk (excluding the Sco-Cen and Cygnus contributions) is shown in Fig. \ref{spec26Al_galaxy}. The centroid and width of the line are consistent with no Doppler shift and/or broadening, but it should be emphasised that this spectrum results from the contributions of many regions along the Galactic plane, each region having specific physical properties like the proper motion (relative to us) due to the Galactic rotation pattern. A study of the spectral characteristics of the 1809\,keV line along the Galactic plane can be found in \citet{Diehl:2006} and \citet{Wang:2008}.

\subsection{The 1173/1332\,keV Galactic emission}
\label{galaxy_iron}

\begin{figure}[!t]
\begin{center}
\includegraphics[width=\columnwidth,height=6.44cm]{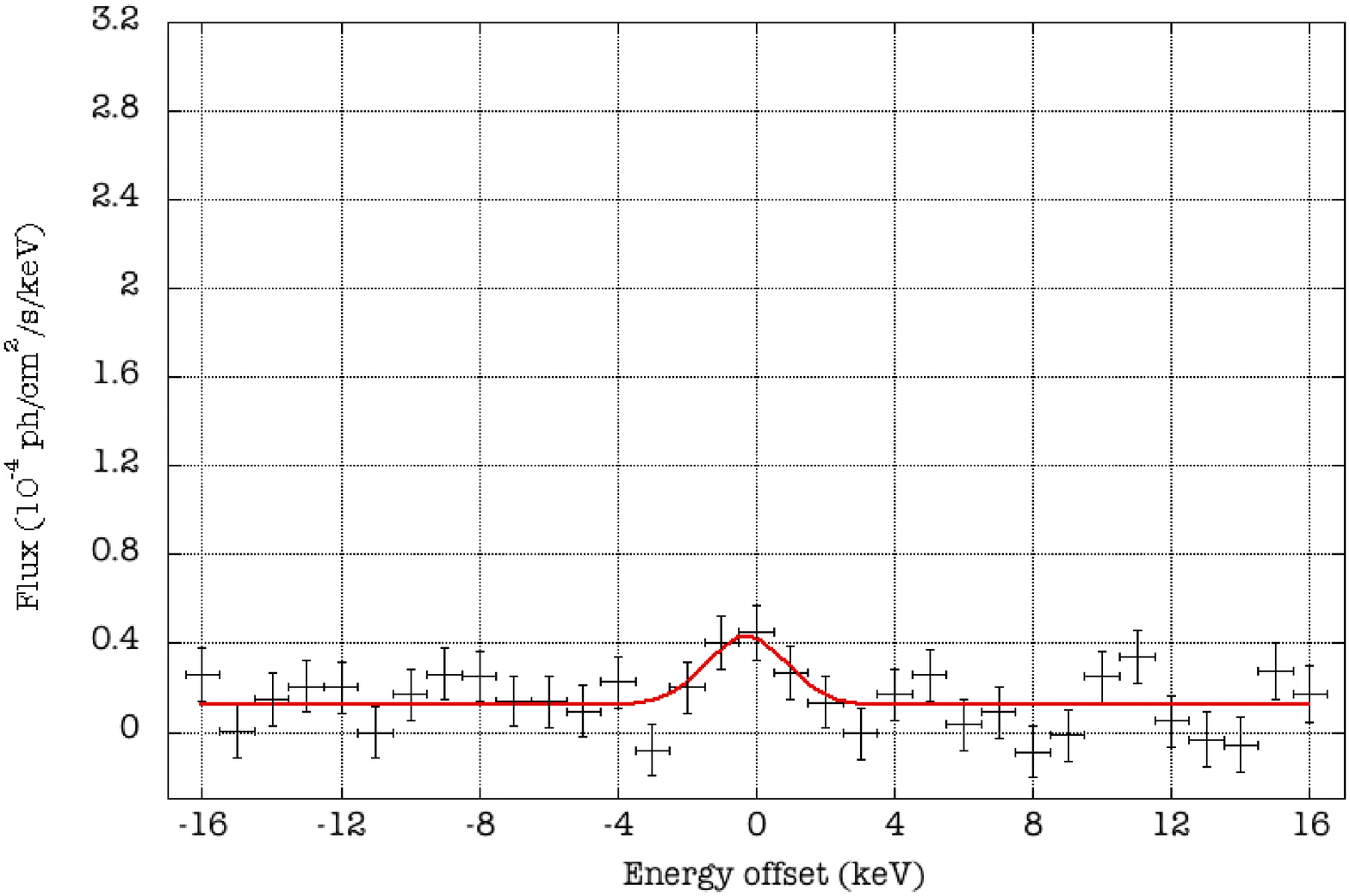}
\caption{Spectrum of the combined 1173/1332\,keV emission from the Galaxy, from about 4 years of INTEGRAL/SPI observations. The red line represents the best Gaussian fit to the data points.}
\label{spec60Fe_galaxy}
\end{center}
\end{figure}

\indent Due to lower intensity, the decay emission of $^{60}$Fe cannot be analysed in as much detail as the emission from $^{26}$Al; in particular, the spatial distribution of the $^{60}$Fe isotope in the Galaxy cannot be estimated, even for a simple model. The only observational constraint accessible to us is the emission flux, especially the flux from the Galactic centre region (where most of the emission and exposure time are concentrated).\\
\indent The decay emission at 1173 and 1332\,keV from the $^{60}$Fe isotope is assumed to follow the distribution of massive stars throughout the Galaxy (it should be remembered here that, in contrast to $^{26}$Al, it has never been demonstrated observationally that massive stars are the dominant source of $^{60}$Fe). As a tracer of the massive stars, we used the infrared emission from the interstellar dust heated by the strong UV luminosity of OB stars, as observed at 240\,$\mu$m by the COBE/DIRBE instrument. Fitting this intensity distribution to the data in the 1150-1190 and 1300-1350\,keV bands (simultaneously to a background model for each energy range) and combining the resulting spectra gives the spectrum of Fig. \ref{spec60Fe_galaxy} for the Galactic $^{60}$Fe decay emission.\\
\indent The line is too weak to allow for a solid spectrometric study, so we focus on photometric considerations. Assuming the  line width is purely instrumental, the flux in the 1173 and 1332\,keV decay lines is (9.1 $\pm$ 2.6) $\times$ 10$^{-5}$\,ph\,cm$^{-2}$\,s$^{-1}$ for the whole Galaxy and (3.5 $\pm$ 1.0) $\times$ 10$^{-5}$\,ph\,cm$^{-2}$\,s$^{-1}$ for the central region (defined as a steradian centred on the Galactic centre). It is worth mentioning that the flux value obtained for the Galactic centre region is very likely more reliable than the one for the whole Galaxy. Indeed, due to the weak emission intensity in the Galactic disk and because of the strongly inhomogeneous exposure of INTEGRAL, the scaling of the DIRBE intensity distribution is driven by the Galactic centre region; as a result, the flux at high longitudes is more a consequence of this "biased" scaling than a solid measurement.

\subsection{Discussion}
\label{galaxy_discu}

\indent The chacteristics of the $^{26}$Al and $^{60}$Fe decay emissions from the Galaxy agree with the results obtained in previous studies using INTEGRAL/SPI or other instruments. The flux values are consistent with all previous measurements. Our analysis of the 1173/1332\,keV lines yielded a flux in the inner Galaxy of (3.5 $\pm$ 1.0) $\times$ 10$^{-5}$\,ph\,cm$^{-2}$\,s$^{-1}$, in agreement with the values of (3.6 $\pm$ 1.4), (3.7 $\pm$ 1.1) and (4.4 $\pm$ 0.9) $\times$ 10$^{-5}$\,ph\,cm$^{-2}$\,s$^{-1}$ produced from RHESSI \citep{Smith:2004} and SPI \citep{Harris:2005,Wang:2007} observations.\\
\indent Regarding the Galactic 1809\,keV emission, the emission intensity we derived is also consistent with previous estimates \citep[see for instance][]{Diehl:2006,Wang:2008}. We found a $^{26}$Al stationary mass of 1.7 $\pm$0.2\,M$_{\odot}$ (excluding the specific Sco-Cen and Cygnus contributions), most of which is confined between 3 and 6\,kpc, at the position of the Giant Molecular Ring. This value, however, is linked to an assumption on the latitudinal extent of the 1809\,keV Galactic emission. This parameter cannot be strongly constrained by the SPI observations due to intrinsic limitations of the coded-mask imaging system (see the detailed explanation given in \ref{cygnus_morpho}). Therefore, we adopted a characteristic value of 90\,pc for the vertical extent of the 1809\,keV emission. This is a "minimal" hypothesis since it is the characteristic scale height of the disk of molecular gas out of which massive stars form; since massive stars are the dominant source of $^{26}$Al, the vertical extent of the 1809\,keV emission cannot be below these 90\,pc. Nevertheless, the ejected $^{26}$Al will diffuse in the ISM, in particular inside the superbubbles blown by the collective winds and SN explosions of the OB associations, and can travel out to distances of a few hundred pc. The 1809\,keV emission is therefore expected to be more extended vertically than 90\,pc. To evaluate the importance of this parameter, we performed the analysis of the 1809\,keV all-sky emission using as a disk height for the Galactic ring model a value of 300\,pc instead of the minimal 90\,pc. The fit to the data (using the same background model) was found to be identical in terms of statistical criterion, but the Galactic mass of $^{26}$Al rose to 2.0 $\pm$0.2\,M$_{\odot}$.\\
\indent Our estimate of the $^{26}$Al stationary mass in the Galaxy is lower than but compatible with, within the error bars, the values put forward by \citet{Diehl:2006} and \citet{Wang:2008}, which are respectively 2.8 $\pm$0.8\,M$_{\odot}$ and 2.7 $\pm$0.7\,M$_{\odot}$. The larger uncertainties associated with these values actually reflect our poor knowledge of the exact sky distribution of the 1809\,keV emission and associated three-dimensional Galactic distribution of the $^{26}$Al isotope. Indeed, these authors have used, as we did in the present work, a model-fitting approach to extract the photometric (and spectrometric) information about the 1809\,keV Galactic emission; in order to evaluate the dependence on the assumed galactic distribution, they have considered a series of plausible models, and the above-cited values reflect the scatter in the inferred $^{26}$Al masses. Our work complements this study in that our concentric ring model has not been considered by \citet{Diehl:2006} and \citet{Wang:2008}, and this model implies a $^{26}$Al mass at the lower end of the mass range they determined. On that point, it is interesting to note that, among the models tested by \citet{Wang:2008}, the best fitting ones are actually associated with the lowest $^{26}$Al mass values, with a meaningful (but not drastic) statistical improvement (Michael Lang 2008, private communication). A major difference in our analysis is also that we separately modelled the Cygnus and Sco-Cen complexes of massive stars. It may be that these peculiar nearby regions, on which there is considerable SPI exposure, somehow drive the fit of some all-sky models to higher underlying $^{26}$Al masses. Ongoing INTEGRAL observations of the Galaxy, and specifically of the Cygnus and Sco-Cen regions, will help refining our perception of the 1809\,keV emission and associated $^{26}$Al distribution.\\
\indent As already said in \ref{cygnus_spectro}, a major result of our all-sky analysis of the 1809\,keV signal is a better estimate of the 1809\,keV decay flux attributable to the Cygnus complex. We indeed disentangled the contribution of the nearby clustered OB population of the Cygnus complex in a coherent way from the contribution of the more diffuse and spatially extended non-clustered population likely to contribute to the emission. Also noteworthy is the non-negligible 1809\,keV flux found from the Sco-Cen region. A similar flux was found from almost the same data set and is further discussed by \citet{Diehl:2009}.\\

\section{Conclusion}
\label{conclusion}

We have analysed about 4 years of data from the INTEGRAL/SPI spectrometer to characterise the $^{26}$Al and $^{60}$Fe decay emission from the Cygnus region in detail, with the purpose of constraining the nucleosynthesis activity of the Cygnus complex of massive stars located between 1.0 and 2.0\,kpc from the Sun.\\
\indent The $^{26}$Al decay emission from Cygnus can be represented as a $3^{\circ} \times 3^{\circ}$ Gaussian centred on $(l,b)=(81^{\circ},-1^{\circ})$, a position consistent with that of the massive Cyg OB2 cluster thought to dominate the energetics and nucleosynthesis of the Cygnus complex. The total 1809\,keV flux from the Cygnus region is $(6.0 \pm 1.0) \times 10^{-5}$ ph\,cm$^{-2}$\,s$^{-1}$ and the flux really attributable to the Cygnus complex, obtained by properly considering the mean Galactic background and foreground contribution, reduces to $(3.9 \pm 1.1) \times 10^{-5}$ ph\,cm$^{-2}$\,s$^{-1}$. This effort to coherently disentangle the contribution of the rather well-identified clustered OB population of the nearby Cygnus complex from the contribution of the more diffuse and less known non-clustered population will allow a more accurate comparison of observations with theoretical expectations based on the most recent stellar models (to be presented in a forthcoming paper).\\
\indent The 1809\,keV line centroid agrees with expectations considering the direction and distance of the Cygnus complex, and the line width is consistent with typical ISM motions. No decay emission from $^{60}$Fe was observed from the Cygnus region and an upper limit of $1.6 \times 10^{-5}$ ph\,cm$^{-2}$\,s$^{-1}$ was derived.\\
\indent In the course of our work, a modelling of the Galactic $^{26}$Al content and decay emission was performed. In agreement with previous results, we have found a total 1809\,keV flux of (8.4 $\pm$ 1.7) $\times$ 10$^{-4}$\,ph\,cm$^{-2}$\,s$^{-1}$. Our model implies a $^{26}$Al stationary mass of 1.7-2.0 $\pm$0.2\,M$_{\odot}$, depending on the ill-constrained vertical extent of the 1809\,keV Galactic emission, and most of this mass appears to be located in the Giant Molecular Ring, between 3-6\,kpc. We also observed a weaker Galactic $^{60}$Fe decay emission with a total flux of (9.1 $\pm$ 2.6) $\times$ 10$^{-5}$\,ph\,cm$^{-2}$\,s$^{-1}$, which reduces to (3.5 $\pm$ 1.0) $\times$ 10$^{-5}$\,ph\,cm$^{-2}$\,s$^{-1}$ in the central region of the Galaxy.

\begin{acknowledgement}
The SPI project was completed under the responsibility and leadership of CNES. We are grateful to the ASI, CEA, CNES, DLR, ESA, INTA, NASA, and OSTC for their support.
\end{acknowledgement}

\bibliographystyle{aa}
\bibliography{/Users/pierrickmartin/Documents/MyPapers/biblio/Cygnus&CygOB2,/Users/pierrickmartin/Documents/MyPapers/biblio/Nucleosynthesis,/Users/pierrickmartin/Documents/MyPapers/biblio/SNobservations,/Users/pierrickmartin/Documents/MyPapers/biblio/SPI,/Users/pierrickmartin/Documents/MyPapers/biblio/26Al&60Fe,/Users/pierrickmartin/Documents/MyPapers/biblio/SNRobservation,/Users/pierrickmartin/Documents/MyPapers/biblio/Galaxyobservations,/Users/pierrickmartin/Documents/MyPapers/biblio/44Ti,/Users/pierrickmartin/Documents/MyPapers/biblio/ScoCen,/Users/pierrickmartin/Documents/MyPapers/biblio/StellarModels}

\end{document}